\documentclass[aps,prc,preprint,showpacs,preprintnumbers,amsmath,amssymb] {revtex4-1}
\usepackage{amsmath}
\usepackage{amssymb}
\usepackage{graphicx}
\usepackage{subfigure}
\usepackage{rotating}
\usepackage{natbib}
\usepackage{txfonts}
\usepackage{color}
\usepackage{bbold}
\def\beq{\begin{equation}}
\def\eeq{\end{equation}}
\def\bea{\begin{eqnarray}}
\def\eea{\end{eqnarray}}

\renewcommand{\vec}[1]{\boldsymbol{#1}}

\newcommand{\La}{\ensuremath{\mathcal L}}

\newcommand*{\figref}[1]{Fig.~\ref{fig:#1}}
\newcommand*{\figlab}[1]{\label{fig:#1}}

\def\eqref#1{Eq.~(\ref{eq:#1})}
\def\eqlab#1{\label{eq:#1}}

\def\be{\begin{equation}}
\def\ee{\end{equation}}
\def\bg{\begin{eqnarray}}
\def\en{\end{eqnarray}}

\long\def\Omit#1{}

\DeclareGraphicsExtensions{ .pdf, .jpg, .eps}
\begin{document}
\title{Production of the $H$ dibaryon via the ($K^-,K^+$) reaction on a $^{12}$C target}
\author{R. Shyam$^1$}
\author{O. Scholten$^2$}
\author{A.W. Thomas$^3$}
\affiliation{$^1$Saha Institute of Nuclear Physics, 1/AF Bidhan Nagar, Kolkata 700064, India
and Physics Department, Indian Institute of Technology, Roorkee, India}
\affiliation{$^2$Kernfysisch Versneller Instituut, University of Groningen, NL-9747 
Groningen, The Netherlands}
\affiliation{$^3$Special Research Centre for the Subatomic Structure of Matter (CSSM)
and ARC Centre of Excellence in Particle Physics at Terascale (CoEPP),School of  Chemistry
and Physics, the University of Adelaide, Adelaide, SA 5005, Australia}

\date{\today}

\begin{abstract}
We study the production of the stable six-quark $H$ dibaryon via the 
$(K^-,K^+)$ reaction on a $^{12}$C target within a covariant effective 
Lagrangian model. The calculations are performed within a factorization 
approximation, in which the full production amplitude is written as a product 
of the amplitudes for the $K^- + p \to K^+ + \Xi^-$ and $\Xi^- + p \to H$ 
processes. The $K^+\Xi^-$ production vertex is described by excitation, 
propagation and decay of $\Lambda$ and $\Sigma$ resonance states in the initial 
collision of a $K^-$ meson with a target proton in the incident channel. The 
parameters of the resonance vertices are taken to be the same as those 
determined previously by describing the available data on total and differential 
cross sections for the $p(K^-, K^+)\Xi^-$ reaction within a similar model. The 
$\Xi^- + p \to H$ fusion process is treated within a quark model where the $H$ 
dibaryon is considered as a stable particle.  For the $K^+$ meson angle fixed at 
0$^\circ$, the $H$ production cross-section is found to be about 2.9 $\mu b/sr$ 
for $H$ mass just below the $\Lambda \Lambda$ threshold at a $K^-$ beam momentum 
of 1.67 GeV/c. This is an order of magnitude larger than the value for this 
quantity  reported earlier in calculations performed on a $^3$He target using a 
different model for the cascade hyperon production. We have also calculated the 
beam momentum dependence of the $H$ production cross section and the energy 
spectrum of the emitted $K^+$ meson.     
\end{abstract}
\maketitle

\section{Introduction}

Within the quark-bag model, the $H$ dibaryon, a six-quark [two up ($u$), two down 
($d$) and two strange ($s$)] state with spin-parity  $J^\pi = 0^+$, and isospin 
$I = 0$,  was predicted to be a stable system with a mass about 80 MeV below 
the $\Lambda \Lambda$ threshold, some 35 years ago~\cite{jaf77}. Later calculations, 
which included the center-of-mass (c.m.)~\cite{liu82} and pionic-cloud~\cite{mul83} 
corrections within this model, predicted this state to be much less bound or even 
unbound. Around the same time calculations performed within a quark cluster model 
also found it unbound~\cite{oka83}. Since then a lot of experimental effort has 
gone into searching for the $H$ dibaryon [see, e.g., the review~\cite{sak00} 
for references up to the year 2000 and Refs.~\cite{yoo07} and~\cite{neh11} for more 
recent investigations done by Japan's National Laboratory for High Energy Physics 
(KEK) and the STAR Collaboration at the BNL Relativistic Heavy ion Collider, 
respectively]. These studies have led to the conclusion that the existence of this 
system as a deeply bound object is highly unlikely. At the same time, the observation 
of the double-$\Lambda$ hypernucleus $^6_{\Lambda \Lambda}$He (NAGARA event) and the 
precise determination of its binding energy at KEK in the experiment E373 \cite{tak01} 
have put a lower limit of 2.224 GeV to the $H$ dibaryon mass at a 90$\%$  confidence 
level, which is just about 6.9 MeV below the $\Lambda \Lambda$ threshold. 

The interest in the $H$ dibaryon has been revived by the recent lattice quantum 
chromodynamics (LQCD) calculations of different groups. The NPLQCD~\cite{bea11a} 
and HAL QCD~\cite{ino11} collaborations have reported that the $H$ particle is 
indeed bound at somewhat larger than physical pion masses. However, extrapolations 
of the calculations of these groups to the physical pion mass region suggest 
\cite{sha11,hai11,bea11b} that this particle is likely to be in either a very 
loosely bound state or an unbound state near the $\Lambda \Lambda$ threshold. In a 
very recent chiral constituent quark model calculation~\cite{car12}, the value 
extracted for the binding energy of the $H$ particle has been found to be compatible 
with the restrictions imposed by the NAGARA event. These results together with the 
previous experiments~\cite{yoo07} that give an upper limit for the cross section of 
the $H$ production in the $^{12}$C($K^-,K^+ \Lambda\Lambda)X$ reaction, have led to a 
proposal to look for this particle in a future experiment~\cite{ahn12} at the Japan 
Proton Accelerator Research Complex (JPARC) using a high-intensity $K^-$ beam. This 
is expected to answer the long standing question about the existence of the 
$H$ dibaryon. Furthermore, with the measurement of the exclusive $\Xi^-$ production 
in the $\gamma p \to K^+ K^+ \Xi^-$ reaction at the Jefferson Laboratory~\cite{guo07}, 
a possibility has been opened for producing the $H$ dibaryon with a photon beam.    

The $(K^-,K^+)$ reaction leads to the transfer of two units of both charge and 
strangeness to the target nucleus. Thus this reaction is one of the most promising 
ways of studying the production of $S = -2$ systems such as $\Xi$ hypernuclei and  
the $H$ dibaryon. Recently, the production of cascade hypernuclei via the 
$(K^-,K^+)$ reaction on nuclear targets, has been investigated within an effective 
Lagrangian model~\cite{shy12,shy13}. This is a new approach, where the $K^+\Xi^-$ 
production vertex is described by excitation, propagation and decay of $\Lambda$ 
and $\Sigma$ resonance intermediate states in the initial collision of the $K^-$ 
meson with a target proton in the incident channel. The $\Xi^-$ hyperon gets 
captured into one of the nuclear orbits leading to the formation of the $\Xi^-$ 
hypernucleus.  In calculations of the  $\Xi^-$ hypernuclear production cross sections, 
one requires the bound state spinors for the proton hole and $\Xi^-$ particle bound 
states. These were  obtained by solving the Dirac equation with vector and scalar 
potential fields having Wood-Saxon shapes. Their depths were fitted to the binding
energies of the respective states. In Ref.~\cite{shy12} bound state spinors obtained 
in the quark-meson coupling (QMC) model~\cite{gui08} were also used. The cross 
sections for the hypernuclear production were found to be quite different from those 
calculated previously in Ref.~\cite{dov83}.

In Ref.~\cite{shy12}, the parameters at the resonance vertices were determined by 
describing the available data on total and differential cross sections for the 
elementary process $^1$H$(K^-, K^+)\Xi^-$ within a similar effective Lagrangian 
model~\cite{shy11,shy12}, where contributions were included from the $s$-channel and 
$u$-channel diagrams, which have as intermediate states $\Lambda$ and $\Sigma$ 
hyperons together with eight of their three-and four-star resonances [$\Lambda(1405)$, 
$\Lambda(1520)$, $\Lambda(1670)$, $\Lambda(1810)$, $\Lambda(1890)$, $\Sigma(1385)$, 
$\Sigma(1670)$ and $\Sigma(1750)$] with masses up to 2 GeV.  It was observed that 
the total cross section of the $^1$H$(K^-, K^+)\Xi^-$ reaction is dominated by the 
contributions from the $\Lambda(1520)$ (with $L_{IJ} = D_{03}$) resonance i
intermediate state. The region for beam momentum ($p_{K^-}$) below 2.0 GeV/c was 
found to be dominated by contributions from the $s$-channel graphs - the $u$-channel 
terms are dominant only in the region $p_{K^-}$ $\le$ 2.5 GeV.  

\begin{figure}[t]
\centering
\includegraphics[width=.50\textwidth]{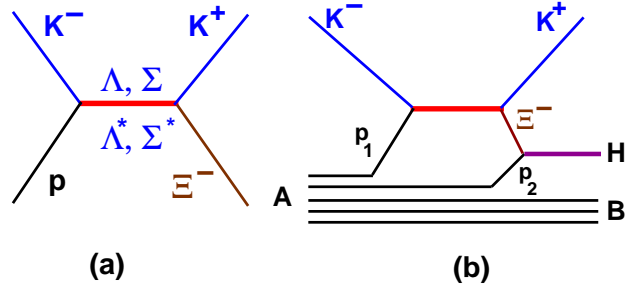}
\caption{(Color online)
Graphical representation of the model used to describe the $^1$H$(K^-, K^+)\Xi^-$ 
[Fig.~1(a)] and $A(K^-, K^+)H B$  reactions [Fig. 1(b)], where $A$ represents
the target nucleus while $B = (A-2)$ the residual nucleus. In Fig.~1(a),  
$\Lambda^*$ and $\Sigma^*$ represent the $\Lambda$ and $\Sigma$ resonance states,
respectively. 
}
\label{fig:Fig1}
\end{figure}

In this paper, we describe the production of the $H$ dibaryon by the $(K^-,K^+)$ 
reaction on a $^{12}$C target using a similar approach. The basic production mechanism 
considered in our work is depicted in Fig.~1, where this reaction proceeds in two 
steps. In the first step the $K^ + \Xi^-$ production takes place by following the 
process as described above [see, Fig. 1(a)], while in the second step the $\Xi^-$ hyperon 
fuses with another proton of the residual nucleus to form the $H$ dibaryon. A similar 
method  was also used earlier in Refs.~\cite{aer82,aer83,aiz92} in calculations of the 
production of this particle via $(K^-,K^+)$. However, there are some differences between 
those calculations and the present work. In our method the amplitude for the $K^- + p_1 
\to K^+ + \Xi^-$ reaction is calculated by employing the same method as that described in 
the previous paragraph. However,  $p_1$ is treated as a bound particle and in the 
calculations of the corresponding amplitude a bound state spinor is used for the initial
proton state ($p_1$). Therefore, the dynamics of the $K^+ \Xi^-$ production is intimately 
related to the wave function of the initial proton bound state [see Fig.~1(b)]. On the 
other hand, in the procedure of Refs.  \cite{aer82,aer83,aiz92}, the initial state is 
described by a product of the target wave function and the amplitude of the $K^- + p \to 
K^+ + \Xi^-$ reaction.  The latter is  determined from a parametrization of the sparsely 
available experimental zero degree differential cross-section for this reaction 
[see Fig.~1(a)]~\cite{aer83}. Furthermore, while the previous calculations were limited to 
a very light $^3$He target, we apply our method to $^{12}$C. This makes it possible to 
compare our cross sections directly to the existing experimental results and to make 
predictions for future measurements. 

As discussed in Ref.~\cite{aer83}, there may be several higher order processes through 
which the $H$ dibaryon production can proceed. They lead to the $H$ dibaryon  via 
$\Lambda \Lambda$ or $\Sigma \Sigma$ fusion. The latter are produced via the following 
reactions: $K^- + p \to \Lambda (\Sigma) + \pi;\, \pi + p \to \Lambda (\Sigma) + K^+$, 
or $K^- + p \to K^+ + \Xi^-; \, \Xi^- + p \to \Lambda \Lambda (\Sigma \Sigma)$.  It is 
shown in Ref.~\cite{aer83} that the contributions of such terms are not expected to be 
large and they can be ignored. Therefore, like these authors, we have also neglected 
such diagrams in our study. 

In Ref.~\cite{aer84}, an alternative scheme of $H$ production has been discussed in which 
a tagged $\Xi^-$ hyperon is first produced on a hydrogen target via the reaction
$K^- + p \to K^+ + \Xi^-$; it is then slowed down by passing through a moderator. After 
moderation, the slow $\Xi^-s$ are captured in a second target into an atomic orbit, and 
the $H$ is subsequently produced via, e.g., the processes, ($\Xi^-$ p)$_{atom} \to 
H + \gamma$, ($\Xi^-$ d)$_{atom} \to H + n$ and ($\Xi^-$$^4$He)$_{atom} \to H + t$. To
ensure that an $H$ with sharply defined mass is indeed produced, the $K^+$ and 
monoenergetic $\gamma$, neutron or triton should be detected in coincidence. These  
authors have estimated the branching ratios ($R$) of the three $H$ formation reactions 
relative to the total decay widths of the $\Xi^-$ atoms. It has been found that $R$ 
has a sizable value (in excess of 0.5) for ($\Xi^- d$) and ($\Xi^-$$^4$He) atoms if 
$H$ mass ($m_H$) is very close to the $\Lambda \Lambda$ threshold. Therefore, these 
processes have their merit for such values of $m_H$. However, these estimates of $R$ 
are based on very poorly known amplitudes for the transitions $\Xi N \to \Xi N, \Lambda
\Lambda, and \Sigma \Sigma$ at low momentum and are strongly dependent on the models 
chosen to calculate them. The virtue of the $A(K^-,K^+)HB$ reaction studied in this 
paper is that $H$ production occurs through a second order process within a 
single nuclear target where all the components of the total amplitude can be calculated
relatively more reliably. Furthermore, unlike the $\Xi^-$ atom method, there are no 
weak decay losses of $\Xi^-$ during moderation to low momentum. 
\section{Formalism}

We have followed the procedure and notations of Ref.~\cite{bjo64} in deriving the 
formulas for the invariant cross section of the $K^- + A \rightarrow 
K^+ + H + B$ reaction, which can be written as (see, e.g., Ref.~\cite{shy99})
\begin{eqnarray}
d\sigma & = & \frac{m_H m_A m_B}{\sqrt{[(p_{K^-} p_{A})^2-m_{K^-}^2-m_A^2]}}
                     \frac{1}{4(2\pi)^5}\delta^4(P_f-P_i)|A_{fi}|^2 \nonumber \\
                 & \times &    \frac{d^3p_{K^+}}{E_{K^+}}\frac{d^3p_B}{E_B}
                     \frac{d^3p_{H}}{E_{H}},
\end{eqnarray}
where $A_{fi}$ represents the total amplitude, $P_i$ and $P_f$ represent the sum of 
all the momenta in the initial and final states, respectively; and $m_H$, $m_A$, and 
$m_B$ represents the masses of $H$ dibaryon, and nuclei $A$ and $B$, respectively. The 
cross sections in the laboratory or c.m. systems can be written from this equation by 
imposing the relevant conditions. Summations over final spin states and average over 
initial spin states are implied in $|A_{fi}|^2$.

Following the factorization approximation of Ref.~\cite{aer83}, the total amplitude
$A_{fi}$ is written as the product of the amplitudes for the processes 
$K^- + p_1 \to K^+ + \Xi^-$ [$M (K^- + p_1 \to K^+ + \Xi^-)$], and $\Xi^- + p_2 \to H$ 
[$F(\Xi^- + p_2 \to H)$] [see Fig. 1(b)]. We write
\begin{eqnarray}
A_{fi} & = & \int \frac{d^4p_1}{(2\pi)^4} \int \frac{d^4p_2}{(2\pi)^4} 
\delta(p_1 + p_{K^-}- p_{K^+} - p_{\Xi^-})
\nonumber \\ & \times & \delta(p_H - p_{\Xi^-} - p_2) 
\Big [\sum_{Y^*}M (K^- + p_1 \to K^+ + \Xi^-) \Big ] 
\nonumber \\ & \times & F(\Xi^- + p_2 \to H),
\end{eqnarray}
where $p_{K^-}$, $p_{K^+}$, and $p_{\Xi^-}$ are the four momenta of the incoming and 
outgoing kaons and the $\Xi^-$ hyperon, respectively. In Eq.~(2) the delta functions 
represent the momentum-energy conservation at various vertices. Some of them can be 
used to reduce the dimensionality of the integrations in this equation. 

The amplitude $M (K^- + p_1 \to K^+ + \Xi^-)$, where the summation is done over all the 
resonance intermediate states $Y^*$ as described above, has been determined by following 
the method discussed in Ref.~\cite{shy12}. The effective Lagrangians, the corresponding 
coupling constants and the form factors for the resonance-kaon-baryon vertices, the 
propagators for the intermediate resonances and the bound state and free-space wave 
functions for the bound proton $p_1$ and the intermediate $\Xi^-$ hyperon, respectively, 
as used in the calculations of the amplitude $M$, are discussed in the following. 
 
The effective Lagrangians for the resonance-kaon-baryon vertices for 
spin-$\frac{1}{2}$ and spin-$\frac{3}{2}$ resonances are taken as
\begin{eqnarray}
{\La}_{KBR_{1/2}} & = & -g_{KBR_{1/2}} \bar{\psi}_{R_{1/2}}
                            [\chi\,{i\Gamma}\, {\varphi_K}+
      \frac{(1-\chi)}{M}\,\Gamma\, \gamma_\mu\,(\partial^\mu \varphi_K)]
                            {\psi}_B,\\
{\La}_{KBR_{3/2}} & = & \frac{g_{KBR_{3/2}}}{m_K} \bar{\psi}_{R_{3/2}}^\mu 
        \partial_\mu \phi_{K} \psi_{B} + \text{h.~c.},
\end{eqnarray}
with $M \,=\,(m_R \,\pm\,m_B)$, where the upper sign corresponds to an even-parity and 
the lower sign to an odd-parity resonance with $B$ representing either a nucleon or a 
$\Xi$ hyperon and $R$ representing a resonance. The spinors ${\psi}_B$ are defined later 
on.  The operator $\Gamma$ is $\gamma_5$ (1) for an even-parity (odd-parity) resonance. 
The parameter $\chi$ controls the admixture of pseudoscalar and pseudovector components. 
The value of this parameter is taken to be 0.5 for the $\Lambda^*$ and $\Sigma^*$ states, 
but 0 for the $\Lambda$ and $\Sigma$ states, implying pure pseudovector couplings for 
the corresponding vertices, in agreement with Refs.~\cite{shy99,shy08}. The  Lagrangian 
for spin-$\frac{3}{2}$, as given by Eq.~(4), corresponds to that of a pure 
Rarita-Schwinger form that has been used in all previous calculations of the hypernuclear
production reactions within a similar effective Lagrangian 
model~\cite{shy08,shy04,shy09,shy10}. The values of the vertex parameters were taken to 
be the same as those given in Ref.~\cite{shy12}.  
 
Similar to Refs.~\cite{shy12,shy11}, we have used the following form factor at various 
vertices, 
\begin{eqnarray}
F_m(s)=\frac{\lambda^4}{\lambda^4+(s-m^2)^2},
\end{eqnarray}
where $m$ is the mass of the propagating particle. The cutoff parameter $\lambda$ 
is taken to be 1.2 GeV, the same as that used in Refs.~\cite{shy12,shy11}.

The two interaction vertices of Fig.~\ref{fig:Fig1} are connected by a
resonance propagator. For the spin-$1/2$ and spin-$3/2$ resonances, the
propagators are given by
\begin{equation}\label{eq:propspin12}
\mathcal{D}_{R_{1/2}} = \frac{{\it i} (\gamma_\mu p^\mu + m_{R_{1/2}})}
{p^2 - (m_{R_{1/2}} - {\it i}\Gamma_{R_{1/2}}/2)^2},
\end{equation}
and
\begin{equation}\label{eq:propspin32}
\mathcal{D}_{R_{3/2}}^{\mu\nu} = - 
 \frac{{\it i} (\gamma_\lambda p^\lambda + m_{R_{3/2}})}
{p^2 - (m_{R_{3/2}} - {\it i}\Gamma_{R_{3/2}}/2)^2} P^{\mu\nu} \;,
\end{equation}
respectively. In Eq.~\ref{eq:propspin32} we have defined
\begin{equation}\label{eq:prop32proj}
P^{\mu\nu} =
 g^{\mu\nu} - \frac{1}{3} \gamma^\mu \gamma^\nu - \frac{2}{3m_{R_{3/2}}^2} 
p^\mu p^\nu + \frac{1}{3m_{R_{3/2}}} \left( p^\mu \gamma^\nu - p^\nu \gamma^\mu \right)
\;.
\end{equation}
In Eqs.~(\ref{eq:propspin12}) and~(\ref{eq:propspin32}), $\Gamma_{R_{1/2}}$
and $\Gamma_{R_{3/2}}$ define the total widths of the corresponding resonances. We 
have ignored any medium modification of the resonance widths while calculating the 
amplitude $M$, because information about such changes is scarce and uncertain.

The bound proton wave function [$\psi (p_1)$] is a four component Dirac spinor, 
which is the solution of the Dirac Equation for a bound state problem in the 
presence of external scalar and vector potential fields. This is written as 
\begin{eqnarray}
\psi(p_1) & = & \delta(p_{10}-E_1)\begin{pmatrix}
                    {f(k_1) {\cal Y}_{\ell 1/2 j}^{m_j} (\hat {p}_1)}\\
                    {-ig(k_1){\cal Y}_{\ell^\prime 1/2 j}^{m_j}
                     (\hat {p}_1)} 
                    \end{pmatrix},
\end{eqnarray} 
In our notation $p_1$’ represents a four-momentum, and $\vec{p_1}$ a three-momentum. 
The magnitude of $\vec {p_1}$ is represented by $k_1$, and its directions are 
represented by $\hat {p}_1$.  $p_{10}$ represents the time like component of momentum 
$p_1$. In Eq.~(9) $f(k_1)$ and $g(k_1)$ are the radial parts of the upper and lower 
components of the spinor $\psi(p_1)$, ${\cal Y}_{\ell 1/2 j}^{m_j}$ are the coupled 
spherical harmonics
\begin{eqnarray}
{\cal Y}_{\ell 1/2 j}^{m_j} & = & <\ell m_\ell 1/2 \mu_i | j m_j>
                                Y_{\ell m_\ell}(\hat {p_1}) \chi_{\mu},
\end{eqnarray}
and $\ell^\prime = 2j - \ell$ with $\ell$ and $j$ being the orbital and total 
angular momenta, respectively. $Y$ represents the spherical harmonics, and  
$\chi_{\mu}$ the spin space wave function of a spin-$\frac{1}{2}$ particle.

We assume that the nucleon bound state has a pure single particle-hole 
configuration with the core remaining inert. To simplify the nuclear structure 
problem, we assume that the initial bound proton ($p_1$) is picked up from 
the $1p_{3/2}$ orbit with a binding energy of 15.96 MeV. Although it is 
straightforward to include also those cases where the participating proton occupies 
both $p$ and $s$ orbits. However, picking a proton from the $s$-state will lead to 
an unstable residual nucleus in the present case (see, eg. Refs.~\cite{sel90,sel88}). 
Treatment of such systems is beyond the scope of this work.  

The free-space spinor for the $\Xi^-$ hyperon is written as 
\begin{eqnarray}
\Psi(p_{\Xi^-}) & = \delta(p_{{\Xi^-}_0}-E_{\Xi^-}) &\sqrt{\frac{E_{\Xi^-} + 
m_{\Xi^-}}{2m_{\Xi^-}}} \begin{pmatrix}
                        \chi_{\mu}\\
              \frac{\sigma \cdot {\vec p}_{\Xi^-}}{E_{\Xi^-} + m_{\Xi^-}} \chi_{\mu}
              \end{pmatrix},
\end{eqnarray}
where $p_{{\Xi^-}_0}$ is the time-like component of the momentum $p_{\Xi^-}$.
Because our calculations are carried out in momentum space, they include all the 
nonlocalities in the production amplitudes that arise from the resonance propagators.

We have used a plane wave approximation to describe the relative motion of  
kaons in the incoming and outgoing channels. However, the distortion effects are 
partially accounted for by introducing a reduction factor of 4 to the overall cross 
sections, as described in Ref.~\cite{dov83,shy12}. It should be mentioned that this 
factor corresponds to absorption effects on only the $K^-$ and $K^+$ wave functions. 
There could still be the distortion effect on the $H-B$ relative motion in the final 
channel which is ignored here. 
 
For calculating the amplitude $F(\Xi^- + p_2 \to H)$, we follow the same procedure
as described in Ref.~\cite{aer83}. In this method, the $H$ dibaryon is treated 
as a bound particle with a mass $m_H$; however its six-quark structure is taken 
into account. This implies that the three-quark internal structures of the $\Xi^-$ 
hyperon and the proton $p_2$ [see Fig. 1(b)] also have to be invoked as the formation 
of $H$ is thought of in terms of the fusion of two three-quark bags ($\Xi^-$ and 
$p_2$).   The amplitude $F$ is calculated by taking the overlap of the internal wave 
functions of $H$, $\Xi^-$ and $p_2$, which are described by a Gaussian approximation 
(see Ref.~\cite{aer82}). The final result for the amplitude $F$ is given by
\begin{eqnarray}{\label{fusion}}
F(\Xi^- + p_2 \to H) & = & \Gamma_0 \Big(\frac{2R_pR_H}{R_H^2 + R_p^2}\Big)^3
\Big(\frac{2R_{\Xi^-}R_H}{R_H^2 + R_{\Xi^-}^2}\Big)^3 
\Big(\frac{2R_H^2}{3\pi}\Big)^{3/4} \nonumber \\ 
& \times & exp\Big[-\frac{R_H^2}{12}({\vec p}_2 - {\vec p}_{\Xi^-})^2\Big],
\end{eqnarray}
where the factor $\Gamma_0$ arises from the color-flavor-spin recoupling as defined 
in Ref.~\cite{aer83}. Its value is $\sqrt {1/20}$.  The values of the oscillator 
parameters, $R_p$, $R_{\Xi^-}$ and $R_H$ have been taken to be 0.83, 0.73 and 
0.95 fm, respectively. The chosen value of $R_p$ reproduces the root mean square (rms) 
radius of the proton, while the value of $R_{\Xi^-}$ comes from the quark-bag model 
relation between the proton bag radius and that of the $\Xi^-$. The bag radius of the 
$H$ dibaryon is about 20$\%$ larger than that of the proton (see, eg. Refs.
\cite{liu82,mul83}). Therefore, we have increased the value of $R_H$ over $R_p$ 
accordingly. In deriving Eq.~\ref{fusion}, the normalizations of the baryon wave 
functions have been made consistent with those of the amplitude $M$.  

\begin{figure}[t]
\centering
\includegraphics[width=.40\textwidth]{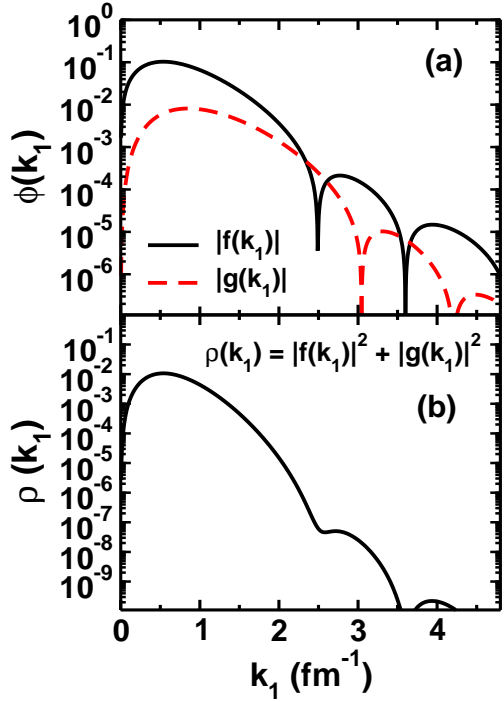}
\caption{(Color online)
(a) Momentum space spinors [$\phi(k_1)$] for the $1p_{3/2}$ nucleon orbit in $^{12}$C.
$f(k_1)$ and $g(k_1)$ are the upper and lower components of the spinor, respectively.
(b) Momentum distribution [$\rho(k_1)$] for the same state calculated with the wave
function shown in panel (a).
}
\label{fig:Fig2}
\end{figure}
\section{Results and discussion}

\subsection{Initial bound state spinors}

The initial bound state spinors in momentum space are obtained by Fourier 
transformation of the corresponding coordinate space spinors, which are the solutions 
of the Dirac equation with potential fields consisting of an attractive scalar part 
($V_s$) and a repulsive vector part ($V_v$), both having  Woods-Saxon shapes. For 
fixed geometry parameters (radius and diffuseness) we search for the depths of these 
potentials to reproduce the binding energy of the respective state. With our choice of 
quantum numbers and the binding energy for the $p_1$ proton state, the resulting depths 
were 382.6 and -472.3 MeV, respectively for the fields $V_v$ and $V_s$, with the 
radius and diffuseness parameters of 0.983 and 0.606 fm, respectively, for both. To 
show the momentum spread of the corresponding spinors, we have displayed in  
Fig.~2(a) and 2(b), the spinors  $|f(k_1)|$ and $|g(k_1)|$ and the momentum 
distribution $\rho (k_1) = |f(k_1)|^2 + |g(k_1)|^2$  as a function of momentum 
$k_1$, respectively. It may be noted that spinors calculated in this way provide 
a good description of the nucleon momentum distribution for the $p$-shell 
nucleons as shown in Ref.~\cite{shy95}. It should further be added here that 
these spinors are the same as those used in Refs.~\cite{shy09,shy12} to 
describe the productions of the $\Lambda$ and $\Xi^-$ hypernuclei via the 
$(\gamma,K^+)$ and $(K^-,K^+)$ reactions, respectively on a $^{12}$C target. 

\begin{figure}[t]
\centering
\includegraphics[width=.50\textwidth]{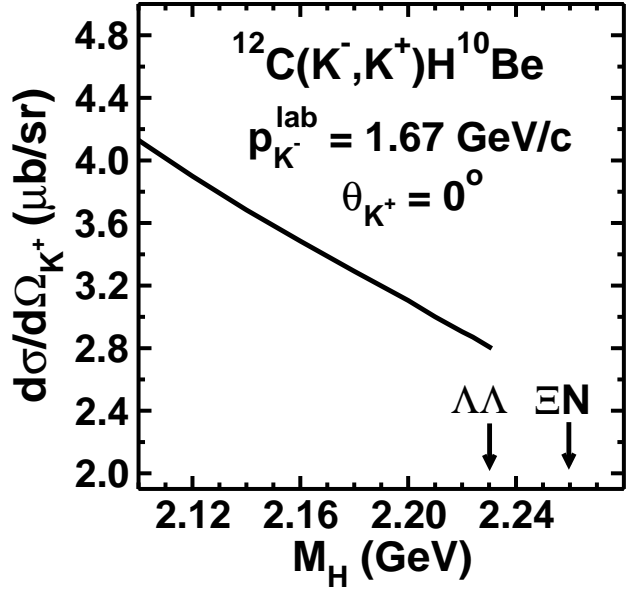}
\caption{
Differential cross section $d\sigma/d\Omega_{K^+}$ at $\theta_{K^+} = 0^\circ$ for 
$H$ production in the $^{12}$C($K^-, K^+)H\,{^{10}}$Be reaction at the beam momentum of
1.67 GeV/$c$, as a function of $H$ dibaryon mass. The $\Lambda \Lambda$ and $\Sigma N$
thresholds are shown by arrows as indicated. 
}
\label{fig:Fig3}
\end{figure}

\subsection{Dibaryon production cross sections}

The method discussed above has been used to study the $^{12}$C($K^-,K^+)H\,{^{10}}$Be 
reaction. In Fig.~3, we show the results for the differential cross section 
$d\sigma/d\Omega_{K^+}$ at the $K^+$ angle of $0^\circ$ and the beam momentum 
($p_{K^-}^{lab}$) of 1.67 GeV/$c$, as a function of the rest mass of the $H$ dibaryon. 
We note that the cross section decreases uniformly as $m_H$ approaches the $\Lambda 
\Lambda$ threshold (indicated by an arrow), where we stopped the calculations because 
our method treats the $H$ as a bound particle. In Ref.~\cite{yoo07}, the upper limit of 
the production cross section of $H$ with a mass range between $\Lambda \Lambda$ and 
$\Sigma N$ thresholds (also indicated in Fig.~2 by an arrow) has been estimated to be 
2.1 $\pm$ 0.6 (stat.) $\pm$ 0.1 (syst.) $\mu b/sr$ at a 90$\%$ confidence level, in a 
measurement of the $^{12}$C($K^-,K^+)\Lambda \Lambda X$ reaction at the beam momentum of 
1.67 GeV/$c$ where the $K^+$ meson was confined mostly in the forward directions. In 
Fig.~3, the cross-section at the $\Lambda \Lambda$ threshold is comparable to this value. 
\begin{figure}[t]
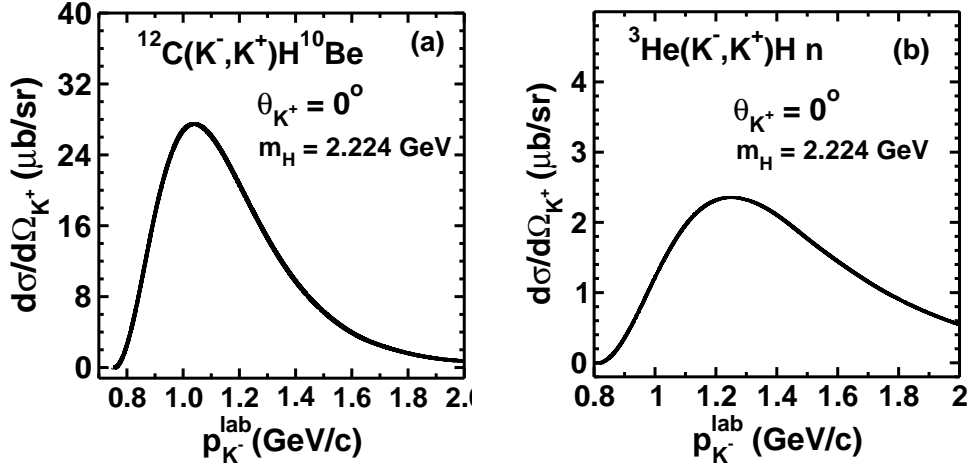

\begin{tabular}{cc}
\includegraphics[scale=0.50]{Fig4a-H.eps} & \hspace{0.30cm}
\includegraphics[scale=0.50]{Fig4b-H.eps}
\end{tabular}
\caption{
Differential cross section $d\sigma/d\Omega_{K^+}$ at $\theta_{K^+} = 0^\circ$ for 
the reactions $^{12}$C($K^-, K^+)H\,{^{10}}$Be (a) and $^{3}$He($K^-, K^+)H\,n$ (b) 
as a function of the laboratory beam momentum $p_{K^-}^{lab}$ corresponding to a $H$ 
dibaryon mass of 2.224 GeV.
}
\label{fig:Fig4}
\end{figure}

In Fig.~4(a), we show the beam momentum dependence of the cross section 
$d\sigma/d\Omega_{K^+}$ for the $^{12}$C($K^-, K^+)H\,{^{10}}$Be reaction. The result  
in this figure, mirrors the beam momentum dependence of the same cross section in the 
$\Xi^-$ hypernuclear production reaction $^{12}$C($K^-,K^+)^{12}\!\!\!{_{\Xi^-}}$Be, 
shown in Refs.~\cite{shy12,shy13}. The cross section peaks near the $p_{K^-}^{lab}$ 
value of about 1.05 GeV/$c$, which is approximately 0.30 GeV/$c$ away from the 
production threshold for this reaction (0.735 GeV/$c$). This is similar to the case of 
the hypernuclear production reactions, as mentioned above, where the threshold is 0.761 
GeV/$c$. We recall that in the case of the zero degree differential cross section for 
the elementary production reaction $p(K^-,K^+)\Xi^-$, the peak also occurs at about 
0.35-0.40 GeV/c above the corresponding production threshold 
(see Refs.~\cite{shy11,shy12}). 

The magnitude of the cross section near the peak position in Fig.~4(a), is about an 
order of magnitude larger than that at the beam momentum of 1.8 GeV/c. This is 
similar to that  seen in Refs.~\cite{shy12,shy13} for the 
$^{12}$C($K^-,K^+)^{12}\!\!\!{_{\Xi^-}}$Be reaction. Moreover, the peak cross 
section of the present reaction is smaller than that of the elementary production 
reaction by roughly a factor of 2. However, it is about an order of magnitude larger 
than that of the $^{12}$C($K^-,K^+)^{12}\!\!\!{_{\Xi^-}}$Be reaction.  This is the 
consequence of the extremely restricted phase space available in the latter case that 
has a two-body final state involving a bound system.

It is evident that in our model the bound state spinors in the initial state are 
calculated within a mean field approximation, therefore, its application to a lighter 
target like $^3$He should be less valid. Nevertheless, in order to have a direct 
comparison with the results shown in Refs.~\cite{aer83,aiz92} for the $^3$He($K^-,K^+)Hn$ 
reaction, we have calculated the beam momentum dependence of the cross section 
$d\sigma/d\Omega_{K^+}$ at $\theta_{K^+} = 0^\circ$ for this reaction within our model. 
The initial bound proton in this case is from the $1s_{1/2}$ orbit with a binding energy 
of 5.49 MeV. The corresponding spinors were determined by a procedure similar to that 
described above using the same geometry parameters. The depths of the fields $V_v$ and 
$V_s$ were 210.41 and -259.76 MeV, respectively. The spinors were normalized to 
reproduce the experimental rms of $^3$He (1.88 fm). Our results for the zero degree 
cross section $d\sigma/d\Omega_{K^+}$ for the corresponding $H$ production reaction are 
shown in Fig.~4b. We note that the peak position of the cross section in this case is at 
a $p_{K^-}^{lab}$ value near 1.20 GeV/$c$. The shift in the peak position to a larger 
$p_{K^-}^{lab}$ as compared to that in Fig.~4(a) can be understood from the fact that 
the threshold for the $^3$He($K^-,K^+)Hn$ reaction ($\approx$ 0.80 GeV/$c$) is larger 
than that of the reaction on a ${12}$C target. 

It should be mentioned here that in Refs.~\cite{aer83} and~\cite{aiz92} the 
corresponding cross section peaks at $p_{K^-}^{lab}$ $\approx$ 1.75 GeV/$c$, which 
coincides exactly with the peak position of the zero degree $d\sigma/d\Omega_{K^+}$ of 
the elementary reaction $^1$H$(K^-,K^+)\Xi^-$ used by these authors in their calculations. 
In this context two points are worth noticing. Firstly, the peak in the elementary 
production cross section calculated in our model occurs at a lower $p_{K^-}^{lab}$ 
(around 1.5 GeV/c) as shown in Refs.~\cite{shy11,shy12,shy13}. Secondly, because in our
calculations of the $A(K^-,K^+)HB$ reactions, the $K^+ \Xi^-$ production amplitudes are
obtained by considering the initial proton as a particle bound in one of the orbits of
the target nucleus, the threshold effects pull back the peak positions in the 
$^3$He($K^-,K^+)Hn$ reaction to a $p_{K^-}^{lab}$ value lower than that of the 
elementary reaction. This effect is not seen in results shown in Refs.
\cite{aer83,aiz92} due to their particular choice of the initial state as discussed in 
the previous section.     

Near $p_{K^-}^{lab} \approx$ 1.75 GeV/$c$,  the magnitude of our cross section is about 
0.95 $\mu b/sr $. This is larger than those of Refs.~\cite{aer83} and~\cite{aiz92}, 
by factors of 2-3 and 1.5-2.0, respectively. This can be attributed to the difference 
in the model used to calculate the $K^+ \Xi^-$ production amplitudes by these authors 
as compared to that of ours.
 
Looking at the results for the $H$ production reactions shown in Figs.~4a and 4b, 
we notice that the ratio ($R_{cs}$) of the magnitudes of the cross sections for the 
$^{12}$C and the $^3$He targets, is greater than one for all the beam momenta. However, 
the point to note is that this ratio is beam momentum dependent. While near peak 
positions the value of $R_{cs}$ is about 10, in the tail region ($p_{K^-}^{lab} > $ 
1.6 GeV/c) it varies between 4 and 2. From Eq.~(1), it is evident that the mass terms 
already make the phase-space factor of the reaction on a $^{12}$C target larger than 
that on $^3$He by nearly an order of magnitude. Of course, there is a strong 
dependence of the product of the rest of the phase-space and the modulus square of the 
amplitudes on various momenta. They combine with the mass terms of the phase space to 
produce the target mass dependence seen in Figs~4(a) and 4(b). Results shown in 
Fig.~4(a) may affect some of the conclusions of Refs.~\cite{ahn96,yam00} where the cross 
sections for the $H$ production via the $(K^-,K^-)$ reaction on a $^{12}$ target have 
been estimated from an extrapolation of the results on a $^3$He target reported in 
Ref.~\cite{aer83}.

\begin{figure}[t]
\centering
\includegraphics[width=.40\textwidth]{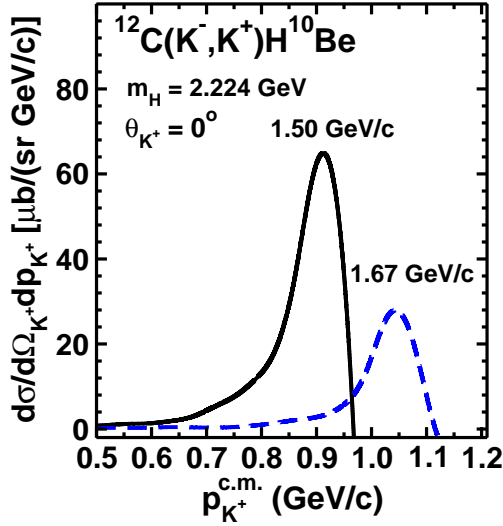}\hspace{0.3cm}
\caption{(Color online)
Double differential cross section $d\sigma/d\Omega_{K^+}dp_{K^+}$ at 
$\theta_{K^+} = 0^\circ$ for the $^{12}$C($K^-, K^+)H\,{^{10}}$Be reaction as a 
function of the c.m. momentum $p_{K^+}^{c.m.}$ of the $K^+$ meson at the 
$p_{K^-}^{lab}$ values of 1.50 GeV/$c$ (solid line) and 1.67 GeV/$c$ (dashed line). 
The $H$ mass is fixed at 2.224 GeV for both the cases. 
}
\label{fig:Fig5}
\end{figure}

In Fig.~5, we show the double differential cross sections 
$d\sigma/d\Omega_{K^+}dp_{K^+}$ for the same reaction as in Fig.~4 at $p_{K^-}^{lab}$ 
values of 1.5 and 1.67 GeV/c and at the $K^+$ c.m. angle $\theta_{K^+} = 0^\circ$, as a 
function of the c.m. momentum of $K^+$ meson ($p_{K^+}^{c.m.}$). The c.m. frame refers 
to that of the $K^- + ^{12}$C system. The $H$ dibaryon mass was taken to be 2.224 GeV 
in both cases. It is seen that these cross sections are peaked very close to the 
kinematically allowed maximum of $p_{K^+}^{c.m.}$ ($p_{K^+}^{c.m.,max}$) and have narrow 
widths of about 90 MeV. The peaking of the cross section near $p_{K^+}^{c.m.,max}$ 
can be understood from the fact that the quark-fusion amplitude is largest for smallest 
values of $\Xi^-$ momenta, which happens for the maximum value of the $K^+$ momentum. 
\begin{figure}[t]
\centering
\includegraphics[width=.40\textwidth]{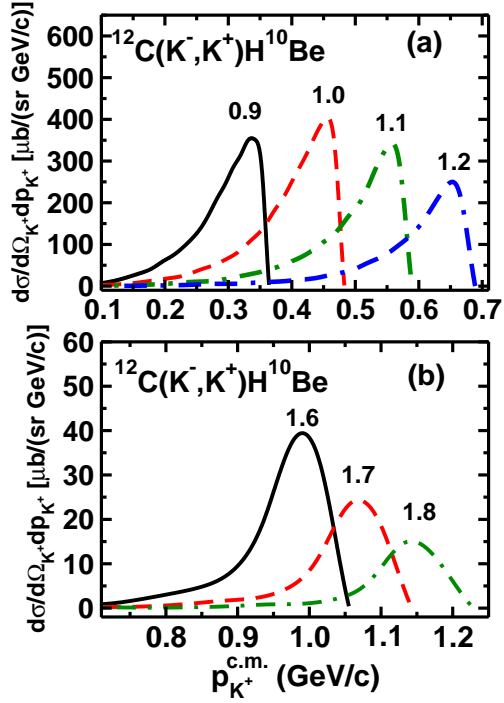}\hspace{0.3cm}
\caption{(color online)
(a) Double differential cross section $d\sigma/d\Omega_{K^+}dp_{K^+}$ at 
$\theta_{K^+} = 0^\circ$ for the $^{12}$C($K^-, K^+)H\,{^{10}}$Be reaction as a 
function of the c.m. $p_{K^+}^{c.m.}$ of the $K^+$ meson for $p_{K^-}^{lab}$ values 
of 0.9 GeV/$c$ (solid line), 1.0 GeV/$c$ (dashed line), 1.1 GeV/$c$ (dashed-dotted line),
and 1.2 GeV/$c$ (dashed-dashed-dotted line). (b) The same as in panel (a) for 
$p_{K^-}^{lab}$ values of 1.6 GeV/$c$ (solid line), 1.7 GeV/$c$ (dashed line) and 
1.8 GeV/$c$ (dashed-dotted line). The $H$ mass is fixed at 2.224 GeV 
for all the cases. 
}
\label{fig:Fig6}
\end{figure}

In Fig.~6,  $d\sigma/d\Omega_{K^+}dp_{K^+}$ is shown for several beam momenta 
($p_{K^-}^{lab}$) for the same reaction as in Fig.~5. It is observed that the 
peaking in the $K^+$ momentum spectra very close to $p_{K^+}^{c.m.,max}$ is found 
in all the cases. As $p_{K^-}^{lab}$ increases $p_{K^+}^{c.m.,max}$ shifts to 
higher values and so does the peak position in the corresponding cross-section.  
However, the widths of the distributions remain unaltered. Nevertheless, at very 
large values of the $p_{K^-}^{lab}$ [in Fig. 6(b)] the distributions tend to become 
more symmetric and the peak positions are at relatively somewhat lower values of 
$p_{K^-}^{c.m.}$ as compared to those at lower $p_{K^-}^{lab}$. This is mainly 
due to more dominant role of the phase space component of the cross section at 
these higher values of $p_{K^-}^{lab}$. The magnitudes of the peak cross sections 
at various values of $p_{K^-}^{lab}$ have a trend that is consistent with that seen 
in Fig.~4.

To understand more clearly the cause for the narrow width of the $K^+$ momentum
spectra, we display in Fig.~7 a decomposition of the cross section 
$d\sigma/d\Omega_{K^+}dp_{K^+}$ into phase-spacer-only (dotted line), $\Xi^-$ 
production-only (dashed line) and quark-fusion-only (solid line) components. The  
$p_{K^-}^{lab}$ is chosen to be 1.67 GeV/$c$ and various curves in this figure are 
normalized to the same maximum value. The three-body phase-space component is broad 
and has a peak at $K^+$ momentum much below the corresponding $p_{K^+}^{c.m.,max}$ (
$\sim$ 1.05 GeV). Therefore, if the shape of the $K^+$ momentum spectrum were 
estimated from the pure phase space, it would have a much more spread-out distribution. 
Although the $\Xi^-$-production-only component is narrower than the phase-space-only 
component, it is still broader than the cross sections shown in Fig.~5. On the other 
hand, the quark-fusion-only component has a narrower width that is similar to that of 
the final cross section. Because of the domination of this component in the cross 
section near the peak position, the shape of the $K^+$ spectrum is governed by it in the 
region of interest. The magnitude of the width of the $K^+$ momentum spectrum obtained 
by us can be understood from simple kinematical arguments as discussed in the Appendix.  
\begin{figure}[t]
\centering
\includegraphics[width=.40\textwidth]{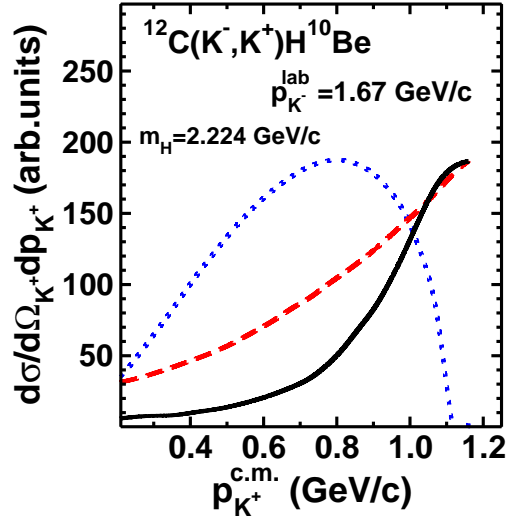}\hspace{0.3cm}
\caption{(Color online)
The phase-space-only (dotted line), $\Xi^-$-production-only (dashed line) and 
$\Xi^- + p \to H$ fusion-only (solid line) components of the differential cross 
section $d\sigma/d\Omega_{K^+}dp_{K^+}$ at $\theta_{K^+} = 0^\circ$ for the 
$^{12}$C($K^-, K^+)H\,{^{10}}$Be reaction as a function of  momentum $p_{K^+}^{c.m.}$ 
at the $p_{K^-}^{lab}$ value of 1.67 GeV/$c$. The $H$ mass is fixed at 2.224 GeV. 
All the curves are normalized to the same peak cross section.
}
\label{fig:Fig7}
\end{figure}

It is important to compare the narrow peak in the $K^+$ spectrum for the $H$ production
reaction with that in the momentum spectrum of the $K^+$ for background processes 
like the $^{12}$C($K^-,K^+)^{10}$Be $\Xi^- p$ reaction. The results of the 
calculations for this type of background reaction on a $^3$He target are discussed in 
Refs.~\cite{aer83,aiz92}. It is shown there that such processes have a relatively 
larger magnitude and a much broader width of the $K^+$ momentum spectrum. Therefore, it 
should be possible to separate the $H$ production from the $\Xi^- p$ background as the 
peak of the $K^+$ spectrum in the latter reaction would be well separated from that for 
$H$ production. There is another potential source of background, namely the possible 
production of two $\Lambda$ hyperons in the interaction of the $\Xi^-$ with proton 
$p_2$. The probability for such processes, initiated by the $\Xi^-$ particle in a nuclear 
bound state, has been calculated in Refs.~\cite{ike94,dov94,yam94}. These calculations 
suggest that the relative probability for the $\Xi^- + p \to \Lambda \Lambda$ reaction 
involving a free $\Xi^-$ particle may not be more than 10-20$\%$.
 
\section{Summary and conclusions}
  
In summary, we have studied the production of the stable six-quark $H$ dibaryon, 
via the $(K^-,K^+)$ reaction on a $^{12}$C target within an effective Lagrangian 
model. We have also made some calculations for this reaction on a $^3$He target. The 
model assumes this reaction to proceed in two steps. In the first step, a $\Xi^-$ 
hyperon and a $K^+$ meson are produced in the initial collision of the $K^-$ meson 
with a proton bound in the target.  In the second step, the $\Xi^-$ hyperon fuses 
with another target proton to produce the $H$ dibaryon. Our method differers from the 
previous calculations of this reaction, which use a similar two step approach, in 
several ways. In our work the $\Xi^- + K^+$ production amplitude has been calculated 
by excitation, propagation and decay of $\Lambda$ and $\Sigma$ hyperon resonance 
intermediate states in the initial collision of the $K^-$ meson with a target proton. 
The vertex parameters (the coupling constants, and the form factors) at the resonance 
vertices have been taken to be the same as those fixed earlier by describing both the 
total and the differential cross sections of the elementary $^1$H$(K^+,K^-)\Xi^-$ 
reaction within a similar model. The same parameters were also used recently in the 
calculations of the $\Xi$ hypernuclei. The bound proton spinors have been  obtained 
by solving the Dirac equation with vector and scalar potential fields having 
Woods-Saxon shapes. Their depths are fitted to the binding energy of the respective 
state. In the previous studies of this reaction, the$\Xi^- + K^+$ production amplitudes 
were obtained from a parametrization of the scantily known experimental differential 
cross sections of the $^1$H$(K^+,K^-)\Xi^-$ reaction at $0^\circ$. Moreover, while we 
have applied our model to compute cross sections for the $(K^-,K^+)$ reaction on both 
$^{12}$C as well as $^3$He targets, in the previous models the numerical calculations 
were limited only to a $^3$He target. 

In our study, the $H$ dibaryon production cross section in the $^{12}$C($K^-, K^+$)
$H\,{^{10}}$Be reaction at the $K^-$ beam momentum of 1.67 GeV/$c$, and for a $H$ mass 
very close to the $\Lambda \Lambda$ threshold, is found to be comparable to the upper 
limit of the $H$ production cross section estimated just above this threshold with 
a 90$\%$ confidence level in a study of the $^{12}$C($K^-,K^+) \Lambda \Lambda X$ 
reaction at the same beam momentum. 

We notice that the differential cross section of the $^{12}$C($K^-, K^+)H\,{^{10}}$Be 
reaction for observing $K^+$ at zero degrees peaks around the beam momentum of 
1.05 GeV/$c$. This mirrors the beam momentum dependence of the corresponding cross 
section in the $\Xi^-$ hypernuclear production reaction 
$^{12}$C($K^-,K^+)^{12}\!\!\!{_{\Xi^-}}$Be.  In case of the $^3$He target the peak 
position shifts to a higher beam momentum of 1.20 GeV/c, because as the threshold of 
the reaction on this target is larger than that on  $^{12}$C. The peak positions in 
these cross sections are above the production thresholds of the corresponding reactions 
by almost the same amount as the position of the maximum is above the corresponding 
threshold in the zero degree differential cross section of the elementary 
$^1$H$(K^-, K^+)\Xi^-$ reaction. In our model, the magnitude of the cross-section 
on a $^{12}$C target is larger than that on  $^3$He by an order of magnitude near the 
respective peak positions. However, in the tail region (for beam momenta larger than 
1.6 GeV/$c$) this difference varies by factors of only 4-2. 

The $K^+$ momentum spectrum has a peak very close to the kinematically allowed 
maximum $K^+$ momentum and its width is narrow (about 90 MeV/$c$). This feature is
independent of the $K^-$ beam momentum.  It is also shown here that a larger $H$
production cross section is expected in experiments performed at beam momenta around 
1 GeV/c. The background process such as the $K^+$ recoiling against the continuum 
$\Xi^- p$ pair~\cite{aer83,aiz92} has a relatively larger magnitude and broader width 
of the $K^+$ momentum spectrum and therefore can be rather cleanly separated from 
the $H$ signal.

\acknowledgments
 
This work was supported by the Dutch Research Foundation (NWO), the Helmholtzzentrum f\"ur 
Schwerionenforschung GmbH (GSI), Darmstadt, the University of Adelaide and the Australian 
Research Council through grant FL0992247(AWT), and Council of Scientific and Industrial
Research (CSIR), India.
\appendix 
\section{Kinematics of H dibaryon production}

To obtain more insight in the kinematical factors that determine the structure 
of the spectrum we have performed a very simple calculation where the focus is on the 
kinematics. We use a convention where the  incoming $K^-$ momentum ($p_{K^-}$), 
is taken in the z-direction like that of the outgoing $K^+$ ($p_{K^+}$). Simplifying the 
reaction dynamics to a minimum the expression for the cross section can be expressed as 
an integral over the longitudinal ($z$) component, $p_r^l$, of the recoil momentum, $p_r$, 
of the $(A-2)$ residual nucleus,
\beq
\sigma_r(p_{K^+})=\int dp_r^l \rho_r(p_r) \,p_r^\perp \;, \eqlab{sigmaR}
\eeq
where $p_r^\perp$ is the perpendicular ($x$) component of the recoil momentum, thus 
$p_r=\sqrt{(p_r^l)^2 + (p_r^\perp)^2}$.  In \eqref{sigmaR} the overlap integral is 
written as
\beq
\rho_r(p_r)=\int d^3{\vec p}_1 \,\rho_1(k_1) \, \rho_2(k_2) \, \rho_H(\Delta p_H) \;,
\eeq
where $\vec{p}_2=\vec{p}_1 - \vec{p}_r$ by momentum conservation. We have followed the 
notations for momenta of bound protons as described after Eq.~(9) in the main text. The 
probability density for a proton with momentum $k_i$ in the nucleus $A$ is parametrized 
as
\beq
\rho_i(k_i)=k_i e^{-(k_i/w_1)^2}/w_1^2 \;,
\eeq
which fits reasonably well the probability distribution given in Fig.~2  for 
$w_1 \approx 0.16$ GeV/$c$. The H dibaryon vertex, is parameterized as
\beq
\rho_H(\Delta p_H)= e^{-(\Delta p_H/w_H)^2}/w_H^2 \;,
\eeq
with  $w_H= 0.73 $ GeV/$c$ where the difference between the cascade momentum and that 
of the second proton, ${\vec p}_2$, is given by
$\Delta p_H =\sqrt{ (p_H^l-2\,p_2^z)^2 + (p_H^\perp-2\,p_2^x)^2 + (-2\,p_2^y)^2 }$.
The total energy of the recoiling nucleus is $\epsilon=[(p_r^l)^2 + (p_r^\perp)^2]/
(2\,B\,m_p) + \beta$ where $\beta$ is the binding energy, which is taken to be 0.008 GeV 
and $B=A-2$. The $H$ dibaryon energy, $E_H $, is obtained from total energy conservation, 
$E_H=E_{K^-}-E_{K^+} + 2\,m_p - \epsilon $. The longitudinal and perpendicular components 
of $\vec{p}_H$ are labeled as $p_H^l $ and $p_H^\perp $, respectively, and are calculated 
from the total momentum conservation, $p_H^l=p_{K^-}-p_{K^+}-p_r^l $, $p_H^\perp= 
-p_r^\perp $. The $H$ dibaryon mass is set at $m_H=2.2$\,GeV. The perpendicular 
component of the recoil is obtained by solving the 'on-shell' condition 
$ m_H^2=E_H^2-p_H^2 $.
\begin{figure}[t]  
\begin{center}
\includegraphics[angle=0,width=0.55\textwidth]{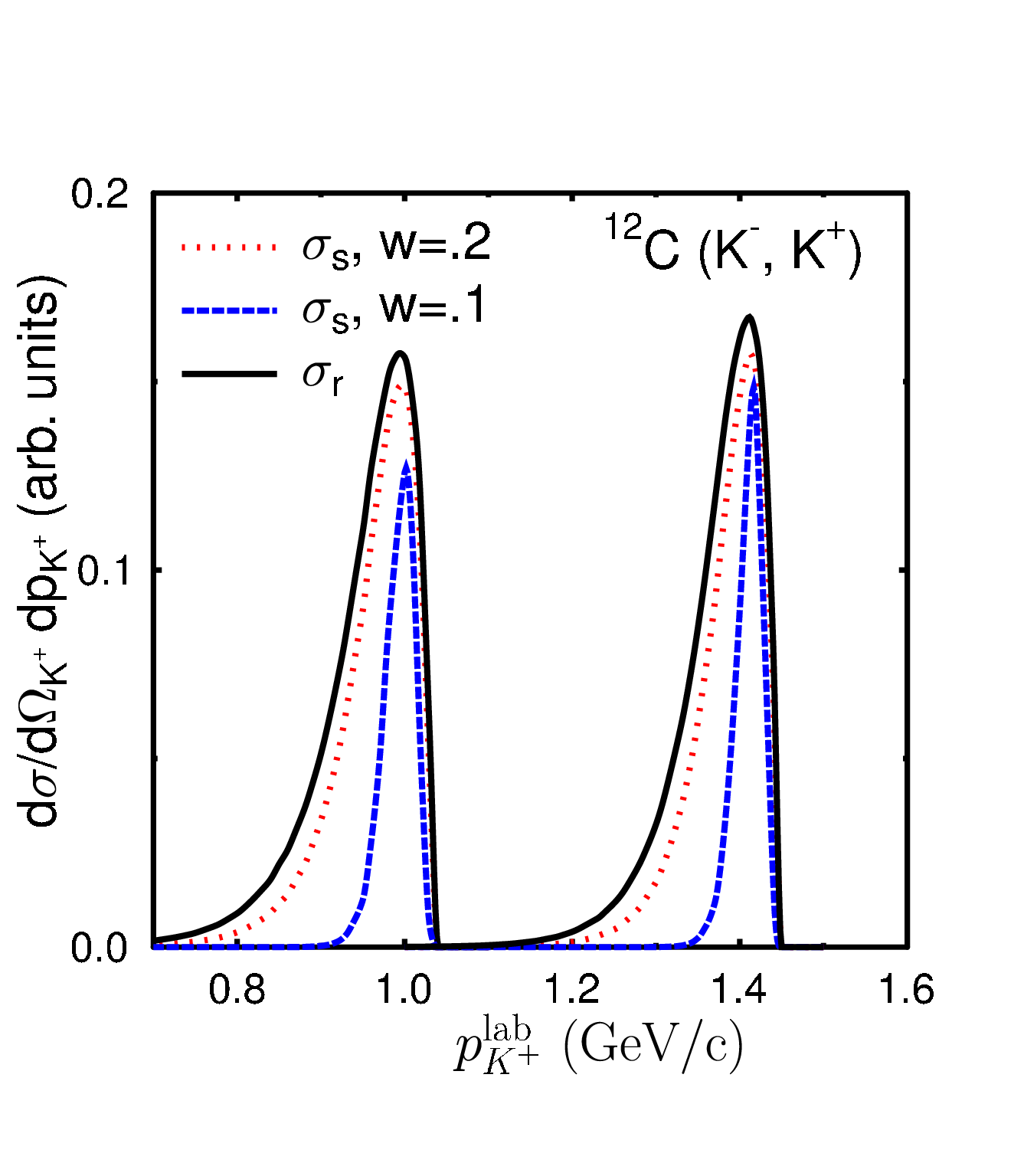}
\caption{(Color online) Cross section calculated with simplified expressions given by 
Eqs.~(A1) and (A5) for $K^+$ angle of 0$^\circ$ as a function of $K^+$ momentum in the
laboratory system at $p_{K^-}^{lab}$ values of  1.4 and 1.8 GeV/$c$. The solid lines 
represent the results obtained with \eqref{sigmaR} while the dashed and dotted lines 
represent those obtained with \eqref{sigmaS} for $W=$\,0.1 and $W=$0.2\,GeV, 
respectively. The absolute magnitudes of the cross section are arbitrary. 
\figlab{DiBarKin}}
\end{center} 
\end{figure}

In \figref{DiBarKin} the cross section obtained from \eqref{sigmaR} is compared with 
that obtained from an even simpler expression
\beq
\sigma_s(p_{K^+})=\int dp_r^l \rho_s(p_r) \,p_r^\perp \;, \eqlab{sigmaS}
\eeq
where the overlap integral is simplified to the extreme as
\beq
\rho_s(p_r)=e^{-(p_r/w)^2}/w^2 \;.
\eeq

In \figref{DiBarKin}, we show the cross sections as a function of $K^+$ momentum in 
laboratory system for $p_{K^-}^{lab}$ values of 1.4 and 1.8 GeV/$c$. The solid 
lines show the results obtained by using \eqref{sigmaR} while dashed and dotted lines
those obtained with \eqref{sigmaS} for $W=0.1$ and $W=0.2$\,GeV, respectively. 
First of all we note that the widths of the cross sections $\sigma_R$ are very close 
to those of the cross sections shown in Figs~5-7. Furthermore, for $W=0.2$\,GeV, the 
cross sections $\sigma_S$ are very close to the cross sections $\sigma_R$. Thus this 
provides an excellent simple parametrization of the cross sections shown in Figs.~5-7. 
The width of the $K^+$ momentum distribution is narrower for smaller values of $W$. 
This is expected to be the case for the bound proton states in heavier targets.

\end{document}